\documentclass[runningheads]{llncs}
\usepackage[T1]{fontenc}
\usepackage{graphicx}
\usepackage[nobreak]{cite}
\usepackage{hyperref}
\usepackage[htt]{hyphenat} 

\usepackage{tabularx}
\usepackage{threeparttable} 
\usepackage{multirow}
\usepackage[usestackEOL]{stackengine} 
\usepackage{cellspace}
\usepackage{xcolor}
\usepackage{colortbl}

\usepackage{pifont}

\usepackage{makecell}

\usepackage{changepage}

\usepackage{graphicx}
\graphicspath{ {./images/} }
\usepackage{pifont}
\usepackage{balance}
\usepackage{soul}
\usepackage{subcaption}
\usepackage{float} 

\usepackage{enumitem}
\usepackage{makecell}
\usepackage{mdframed}

\usepackage{mathtools}
\usepackage{booktabs}
\usepackage{amsmath}

\usepackage{bbm}

\usepackage{algorithm}
\usepackage{algorithmicx}
\usepackage{algpseudocode}

\newcommand{\hi}[1]{\textcolor{black}{#1}}

\newcommand{\shrink}{\vspace*{-.9\baselineskip}}

\newcommand{\cw}[1]{\textcolor{black}{\ul{\textbf{#1}}}}

\usepackage{xspace}
\newcommand{\dataset}{\textsc{WildClaims}\xspace}

\usepackage[most]{tcolorbox}
\usepackage{caption}
\usepackage{lipsum}

\begin{document}
\title{\dataset: Conversational Information Access\\in the Wild(Chat)}
%
%
\author{Hideaki Joko\inst{1}$^*$ 
\and
Shakiba Amirshahi\inst{2}$^*$
\and \\
Charles L. A. Clarke\inst{2}
\and
Faegheh~Hasibi\inst{1} 
}
\authorrunning{H. Joko, S. Amirshahi, C. L. A. Clarke, and F. Hasibi}
\institute{
Radboud University, The Netherlands\\
\email{\{hideaki.joko,faegheh.hasibi\}@ru.nl}
\and
University of Waterloo, Canada\\
\email{shakiba.amirshahi@uwaterloo.ca, claclark@gmail.com}
}
\maketitle              
\begingroup
\renewcommand\thefootnote{}\footnotetext{$^*$ The authors contributed equally.}
\endgroup
\begin{abstract}
\setlength{\parindent}{1.5em}
\noindent The rapid advancement of Large Language Models (LLMs) has transformed conversational systems into practical tools used by millions.
However, the nature and necessity of information retrieval in real-world conversations remain largely unexplored, as research has focused predominantly on traditional, explicit information access conversations.
The central question is: What does real-world conversational information access look like?
To this end, we first conduct an observational study on the WildChat dataset, large-scale user-ChatGPT conversations, finding that users' access to information occurs implicitly as check-worthy factual assertions made by the system, even when the conversation's primary intent is non-informational, such as creative writing.

To enable the systematic study of this phenomenon, we release the \textbf{\dataset} dataset, a novel resource consisting of 121,905 extracted factual claims from 7,587 utterances in 3,000 WildChat conversations, each annotated for check-worthiness.
Our preliminary analysis of this resource reveals that conservatively 18\% to 51\% of conversations contain check-worthy assertions, depending on the methods employed, and less conservatively, as many as 76\% may contain such assertions.
This high prevalence underscores the importance of moving beyond the traditional understanding of explicit information access, to address the implicit information access that arises in real-world user-system conversations.

\vspace{0.5em}
\hspace{3.3em}\includegraphics[width=1.25em,height=1.25em]{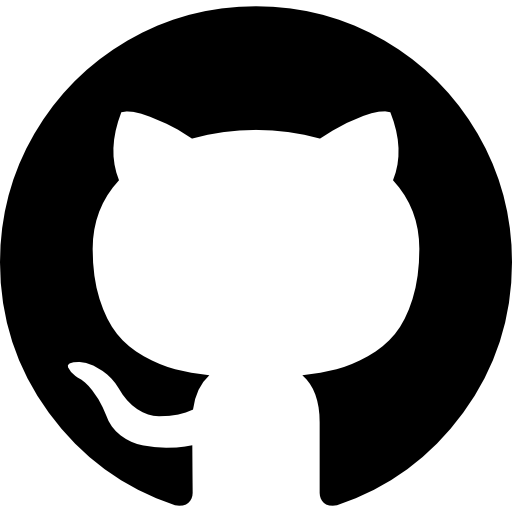}\hspace{.3em}
\parbox[c]{\columnwidth}
{
    \vspace{-.55em}
    \href{https://github.com/shakibaam/wildclaims}{\nolinkurl{https://github.com/shakibaam/wildclaims}}
}
\vspace{-1.4em}


\end{abstract}

\section{Introduction}

The emergence of large language models (LLMs) has drastically changed the way humans interact with machines, leading to a surge in the use of conversational agents for various tasks, including information access~\cite{Trippas:2024:URA,Chatterji:2025:HPU}.
However, the nature of these interactions is not well understood, particularly in terms of how users learn information from these systems.
Understanding the nature of user-system interaction and user behavior is crucial for improving information access systems, building user simulators, and ultimately evaluating these systems~\cite{Zhang:2020:ECR,Balog:2023:USE,Zhang:2024:ULL,Bernard:2025:CRS,Joko:2025:FACE,smucker2016modeling}.

Current literature on conversational information access builds on assumptions about information interaction in chat settings and defines a set of attributes for such systems~\cite{Zamani:2023:CIS,Aliannejadi:2023:iKAT,Dalton:2019:CAT,Radlinski:2017:TFC,Trippas:2025:RFS,Anand:2019:CS,Joko:2024:DPL,Bernard:2025:CRS}.
These assumptions, along with related research into Retrieval-Augmented Generation (RAG)~\cite{Pradeep:2024:INE,Pradeep:2025:RRR,singal-etal-2024-evidence,chen2024benchmarking,siriwardhana2023improving,Amirshahi:2025:ERR} and fact-checking of LLMs~\cite{Tang:2024:MEF,Iqbal:2024:OUF,Tang2024,wei2024long,zhao2023felm,wang2024assessing}, mainly focus on explicit information seeking/access cases, where users' information need is clearly expressed, e.g., ``What is the capital of Canada?'' or ``Which university is better?'' 
Based on these definitions, the research community has built various datasets and shared tasks, focusing on different aspects of information access~\cite {Aliannejadi:2023:iKAT,Dalton:2020:CAT,Dalton:2019:CAT,Pradeep:2025:RRR}; see e.g., a sample from the TREC CAsT dataset in Table~\ref{tab:trec_cast_example}.
However, it remains an open question whether real-world conversational information access, where knowledge is exchanged between humans and systems, actually resembles these examples. 

    \begin{table}[t]
    \caption{Traditional example of conversational information access from TREC CAsT 2019~\cite{Dalton:2019:CAT}.
        Each utterance demonstrates an explicit request for information, which is quite distinct from what we observe in real user-system conversations in the WildChat dataset.}
    \label{tab:trec_cast_example}
    \centering
\begin{tabular}{ll}
\hline
\multicolumn{2}{l}{\textbf{Title}: Uranus and Neptune} \\
\multicolumn{2}{l}{\textbf{Description}: Information about Uranus and Neptune.} \\ \hline
\textbf{Turn} & \textbf{Conversation Utterances} \\ \hline
1 &	Describe Uranus. \\
2 &	What makes it so unusual? \\
3 &	Tell me about its orbit. \\
4 &	Why is it tilted? \\
5 &	How is its rotation different from other planets? \\
6 &	What is peculiar about its seasons? \\
7 &	Are there any other planets similar to it? \\
8 &	Describe the characteristics of Neptune. \\
9 &	Why is it important to our solar system? \\
10 &	How are these two planets similar to each other? \\
11 &	Can life exist on either of them? \\
\hline
\end{tabular}
\shrink
\end{table}

The introduction and widespread adoption of LLMs, such as ChatGPT, has opened up new opportunities for the research community to study how users interact with chat systems to fulfill their information needs. The recently released WildChat dataset~\cite{Zhao:2025:WCI}, comprising over one million real user-ChatGPT conversations, offers a rich resource for investigating and modeling information interaction in conversational settings.
With this dataset, researchers can now explore a fundamental question in the information access systems: \textit{\textbf{RQ1}: ``What does real-world conversational information access look like?''}

\begin{figure}[H]
  \centering
  
  \begin{subfigure}{\linewidth}
    \fontsize{8pt}{9.5pt}\selectfont
    \centering
    \begin{tcolorbox}[colback=gray!5!white, colframe=black!75!black, width=1.0\linewidth, arc=2.5mm, boxrule=1.0pt, left=6pt, right=6pt, top=4pt, bottom=4pt]

      \textbf{User:}
      Write a fake lawsuit for 2 dollars, the incident in Ontario, i was punched while sitting on a public park by a man.

      \vspace{1em}
      \textbf{System:} 
      SUPERIOR COURT OF ONTARIO.
      CLAIMANT: John/Jane Doe.
      DEFENDANT: Mark/Mary Public.
      \cw{Proceeding Commenced under the Torture Act, RSO 1990, c T.21}. [...]

    \end{tcolorbox}
    \shrink
    \captionsetup{width=0.7\linewidth}
    \caption{A legal task response fabricating a non-existent law.}
  \end{subfigure}

  \vspace{0.8em}

  \begin{subfigure}{\linewidth}
    \fontsize{8pt}{9.5pt}\selectfont
    \centering
    \begin{tcolorbox}[colback=gray!5!white, colframe=black!75!black, width=1.0\linewidth, arc=2.5mm, boxrule=1.0pt, left=6pt, right=6pt, top=4pt, bottom=4pt]

      \textbf{User:}
      here's my upwork profile link : \nolinkurl{https://www.upwork.com/freelancers/abcde} please tell me how can i more improve my profile

      \vspace{1em}
      \textbf{System:} 
      Consider the following optimized profile overview: [...]
      ``My expertise in lead generation includes \cw{market research using advanced tools like LinkedIn Sales Navigator and Hunter.io}.
      [...]
      I offer a solid understanding of \cw{CRM tools such as Salesforce, HubSpot, and Zoho, effectively managing customer data}.''

    \end{tcolorbox}
    \shrink
    \captionsetup{width=0.7\linewidth}
    \caption{An editing task response containing factual implications about software capabilities that require verification.}
  \end{subfigure}
  
  \vspace{0.8em}

  \begin{subfigure}{\linewidth}
    \fontsize{8pt}{9.5pt}\selectfont
    \centering
    \begin{tcolorbox}[colback=gray!5!white, colframe=black!75!black, width=1.0\linewidth, arc=2.5mm, boxrule=1.0pt, left=6pt, right=6pt, top=4pt, bottom=4pt]

      \textbf{User:}
      If English knights of the 14th century went to Japan, how would Both knight and Samurai try to use this encounter as political leverage, realistically, taking into account what we know of Both 14th century England and Japan.
      [...] In depth, what would both think of each other's customs?

      \vspace{1em}
      \textbf{System:}
      Here are some aspects of each society that might stand out to the other.
      \cw{1. Bushido Code: The unwritten samurai code of conduct, known as Bushido}, emphasizing honor, loyalty, and courage [...]

      \vspace{1em}
      \textbf{User:}
      In depth, 
      What if one of the knights fell for a Japanese Lady?

      \vspace{1em}
      \textbf{System:} 
      [...] the knight might need to learn how to \cw{comprehend the intricacies of Japanese bowing etiquette} while [...]

    \end{tcolorbox}
    \shrink
    \captionsetup{width=0.7\linewidth}
    \caption{A creative writing task response claiming the 14th-century existence of a ``Bushido code'' and ``Japanese bowing etiquette.''}
  \end{subfigure}
  
\caption{Examples of conversations from the WildChat dataset, demonstrating how various tasks can generate responses with factual implications that require information retrieval and verification. None of these tasks are explicitly information seeking.}
  \label{fig:examples}
\end{figure}


We address this question by studying the WildChat dataset and come to the realization that the necessity for information retrieval is not always in explicit users' information need, but often implicitly in the form of check-worthy factual assertions made by the system; see example conversations in Figure~\ref{fig:examples}. This implies that the existing definitions of information access systems are limited and do not provide a full picture where retrieval is needed.

To enable the study of this phenomenon, we present the \textbf{\dataset} dataset, a novel dataset of 121,905 factual claims extracted from 7,587 utterances across 3,000 WildChat conversations using existing fact extraction methods~\cite{Huo:2023:RSE,Song:2024:VSE}, with check-worthy claims subsequently automatically identified and manually validated on 200 factual claims~\cite{Hassan:2015:DCF,Majer:2024:CCD}. With this resource, we can now quantify this phenomenon: \textit{\textbf{RQ2}: How prevalent is conversational information access in real user-system interactions?}




Our preliminary prevalence analysis conservatively estimates that, depending on the method, 18\%--51\% of conversations contain implicit or explicit factual assertions requiring verification.
Using check-worthiness methods that are fairly correlated with human annotations, this prevalence can be up to 32\%--76\%.
Interestingly, our resource reveals that many of these conversations do not explicitly request information, but rather arise in the context of non-information-access tasks, such as creative writing or editing tasks.

\medskip \noindent \textbf{Key contributions} and findings of this paper are as follows:

\begin{itemize}
   \item We release \textbf{\dataset}, a novel resource containing 121,905 factual claims from 7,587 utterances from 3,000 real-world conversations, each annotated for check-worthiness, together with 200 manually annotated claims for validation. This dataset enables new research into the implicit forms of information access that occur in real-world conversational settings.
    
      \item Enabled by this dataset, we analyze real-world information access in human-ChatGPT interactions, showing that it extends beyond explicit requests, with implicit factual assertions in system responses being prevalent.
    
    \item We provide a modified and more realistic definition of conversational information access systems, which informs the design and development of both systems and user simulators.

\end{itemize}

\section{Information Access in the Wild}
\label{sec:wild}

In this section, we aim to see \textit{what real-world conversational information access looks like (RQ1).}
We first show real user-system interactions from the WildChat dataset~\cite{Zhao:2025:WCI} that highlight counter-examples less considered in current conversational information access literature.
We then review the existing definitions of conversational information access and provide a more general, clarified definition of conversational information access that reflects real user-system interactions.


\subsection{Real-World Conversation Examples}
\label{sec:wild:dataset}

WildChat is a corpus of over one million real user-ChatGPT conversations, approximating authentic user interactions with an information access system.
Figure~\ref{fig:examples} shows real-world examples from this dataset.
While these examples have different user intents (profile editing, legal writing, and creative writing), they all contain factual claims requiring verification against external corpora, such as legal acts, tool capabilities, and historical practices in feudal Japan.
These examples highlight that in real-world interactions, knowledge transfer from the system to the user occurs across various contexts beyond explicit information-seeking cases, including situations seemingly unrelated to information seeking.

\subsection{Theory vs. Reality}
\label{sec:wild:theory}

Having observed these examples, we now look back at what the literature considers conversational information seeking and access, highlight its limitations, and propose a clarified definition of conversational information access that better reflects the reality of human–system interactions.


\medskip \noindent \textbf{Conversational information seeking} is actively studied in the community for more than a decade~\cite{Balog:2021:CAI,mo2024survey,dalton2022conversational,deldjoo2021towards,zamani2020macaw,lajewska2024explainability,Shiga:2017:MIN} with multiple shared tasks and data sets available,
such as the TREC Conversational Assistance Track (CAsT)~\cite{Dalton:2019:CAT,Dalton:2020:CAT,Dalton:2021:CAT,Owoicho:2022:CAT} and the TREC Interactive Knowledge Access Task (iKAT)~\cite{Aliannejadi:2023:iKAT,Aliannejadi:2024:iKAT}.
One of the most widely accepted definitions of conversational information seeking that is adopted by existing benchmarks and datasets~\cite{Aliannejadi:2023:iKAT,Dalton:2019:CAT,Dalton:2020:CAT,Dalton:2021:CAT,Aliannejadi:2024:iKAT,Owoicho:2022:CAT} is from Zamani et al.~\cite{Zamani:2023:CIS}: 
\begin{adjustwidth}{2.5em}{2.5em}
  \vspace{0.5em}
  \textit{``Information seeking conversation is a conversation in which the goal of information exchange is satisfying the information needs of one or more participants.''}~\cite{Zamani:2023:CIS}
  \vspace{0.5em}
\end{adjustwidth}

Observing that recommendations and information seeking, including QA, are commonly treated as two separate types of systems, Balog et al.~\cite{Balog:2021:CAI} calls for a need for seamless integration of these systems, proposing a more unified view as \textbf{conversational information access}, which is defined as:
\begin{adjustwidth}{2.5em}{2.5em}
  \vspace{0.5em}
  \textit{``(Conversational information access systems are) a subset of conversational AI systems that specifically aim at a task-oriented sequence of exchanges to support multiple user goals, including search, recommendation and exploratory information gathering, that require multi-step interactions over possibly multiple modalities.''}~\cite{Balog:2021:CAI}
  \vspace{0.5em}
\end{adjustwidth}

These definitions, while focusing on information access and gathering, do not reflect the variety of interactions between humans and systems, nor do they recognize the potential for hallucination; something that automatic systems, in the past, were not capable of.
Conversational information access, however, \hi{involves users' access to information via factual assertions}, even when users do not perceive themselves as explicitly accessing information or seeking recommendations.


\medskip \noindent \textbf{Clarified Definition of Conversational Information Access.}
Building on previous definitions and our observations from real-world user-system conversations from the WildChat dataset, we propose a clarified definition of conversational information access that reflects the reality of human-system interactions.

\begin{figure}[ht]
  \centering
  \begin{tcolorbox}[
    enhanced,
    colback=blue!5!white,
    colframe=blue!75!black,
    boxrule=0.8pt,
    arc=3pt,
    auto outer arc,
    width=\columnwidth,
  ]
    \textbf{\textit{Conversational information access (in the wild) is a process by which knowledge is transferred to the user to satisfy their needs, regardless of their explicit or implicit information access goals, where such knowledge should be validated and factually accurate.}}

  \end{tcolorbox}
\end{figure}

\section{\dataset Dataset}
\label{sec:dataset}

To enable further research on how users access factual information in real-world user-system conversations, we present the \dataset dataset.
Through this dataset, we aim to answer the second research question: \textit{RQ2: How prevalent is conversational information access in real-world user-system interactions?}
We employ a two-step approach to construct this dataset.
First, using existing automatic methods, we extract a set of factual claims from each system utterance.
Then, knowing that not all factual claims are check-worthy, we automatically annotate and manually validate the extracted factual claims to identify check-worthy factual claims.

\begin{table}[t]
  \caption{Statistics of the \dataset dataset. CW represents check-worthiness.}
  \label{tab:wildclaims_dataset_statistics}
  \centering
  \small
  \setlength{\tabcolsep}{7pt}

  \begin{tabular}{lr|lr}
    \hline
    \multicolumn{2}{c|}{\textbf{General statistics}} & \multicolumn{2}{c}{\textbf{Task category distribution}} \\ \hline
    \# Conversations                    & 3,000           & Information seeking  & 33.5\%  \\
    Single/multi-turn ratio             & 57\% : 43\% & Creative Writing     & 18.8\%  \\
    \# Utterances                       & 15,174          & Editing              & 15.8\%  \\
    \# System utterances                & 7,587           & Reasoning            & 9.9\%   \\
    Avg. utterances per conversation   & 2.52            & Role playing         & 3.1\%   \\
    Avg. words per user utterance      & 95.70           & Planning             & 2.5\%   \\
    Avg. words per system utterance    & 219.24          & Brainstorming        & 2.3\%   \\
    \# Total extracted factual claims   & 121,905         & Advice seeking       & 1.2\%   \\
    \# Automatic CW annotations         & 243,810         & Data Analysis        & 0.4\%   \\
    \# Manual CW annotations            & 200             & Others               & 12.6\% \\ \hline
  \end{tabular}
\end{table}

\subsection{Conversation Selection}

To collect factual claim and check-worthiness annotations on real-world user-system conversations, we first select a subset from WildChat~\cite{Zhao:2025:WCI} as in Section~\ref{sec:wild}.

\medskip \noindent\textbf{Preprocessing.}
We first exclude non-English conversations and those focused on math or coding, as these domains require external tools beyond the scope of our analysis.
Language filtering is done by using the provided labels in the dataset, while math and coding conversations are filtered out using GPT-4.1-mini with the prompt from Zhang et al.~\cite{Zhang:2024:UIC}.
The preprocessing shows that 478,498 of WildChat conversations are in English, and 79.6\% of these are identified as neither math nor coding related.
From these conversations, we randomly sample 3,000 conversations for analysis and resource construction. 

\medskip \noindent \textbf{Task Category Classification.}
To enable analysis of user task distribution, we classify each user utterance using GPT-4.1 following the categories and their definitions from Lin et al.~\cite{Lin:2025:WBL}.

\subsection{Factual Claim Extraction}
\label{sec:claim_ext:experiments}

To study how knowledge access about factual information occurs from the system to the user, we extract factual claims from system utterances.
Formally, for each system utterance $u_s$, we extract a set of factual claims $\mathbf{A} = \{ a_1, a_2, \dots, a_n \}$, where $n$ is the number of factual claims identified in $u_s$. The set $\mathbf{A}$ is obtained via a factual claim extraction method $F$, i.e., $\mathbf{A} = F(u_s, h)$ where $h$ is the conversation history up to current utterance $u_s$.

Some of the factual statements need decontextualization to be self-contained. For example, the claim ``It was released in 2010'' is ambiguous without knowing from the conversation that ``It'' refers to ``The first iPad.''
With this in mind, we select two methods from the various existing approaches~\cite{Huo:2023:RSE,Song:2024:VSE,Min:2023:FFA,hassan2017toward,Kamoi:2023:WRE} that support the generation of self-contained factual claims.
Specifically, we select the methods by Huo et al.~\cite{Huo:2023:RSE} and Song et al.~\cite{Song:2024:VSE} for factual claim extraction, referred to as $F_{Huo}$ and $F_{Song}$, respectively.
The former, $F_{Huo}$, extracts self-contained factual claims by simply prompting an LLM, while the latter, $F_{Song}$, focuses on extracting only verifiable claims.

\medskip \noindent \textbf{Setup.}
We use GPT-4.1 as their backbone model and follow the original implementation and prompting, with minor adjustments to accommodate conversation history. This allows for accurate generation of contextualized facts in multi-turn conversations. All prompts used are available in our repository.

\subsection{Check-Worthiness Classification}
\label{sec:cw:experiments}

Not every factual assertion merits verification, as assertions like ``Cows drink water'' or ``Water is wet'' illustrate; thus, we perform both manual and automatic annotations to identify claims that are worth checking.
Before proceeding, we outline two definitions that will be used throughout the remainder of this paper.

\begin{itemize}
    \item \textbf{Check-worthy Claim.} Check-worthy factual claims refer to factual assertions that are worth fact-check\-ing~\cite{Shaar:2021:OCC,hassan2017toward,arslan2020benchmark}. This work defines a \textit{check-worthy factual claim} as a factual assertion that is worth checking and deemed necessary to be verified through external sources.
    \item \textbf{Check-worthy Utterance/Conversation.} We define a \textit{check-worthy utterance} and \textit{check-worthy conversation} as an utterance and a conversation that contains at least one check-worthy factual claim, respectively.
\end{itemize}

\medskip \noindent \textbf{Manual Check-Worthiness Validation.}
To establish a basis for verifying check-worthiness that can also evaluate automatic methods, the paper authors first manually annotate check-worthiness on the extracted factual claims.
We randomly select 100 factual claims from each method ($F_{Huo}$ and $F_{Song}$) ensuring that each fact comes from a distinct conversation (i.e., 100 distinct conversations for each method), resulting in a total of 200 factual claims.
This number is chosen to ensure high annotation quality while maintaining a reasonable quantity, based on the existing literature~\cite{Joko:2024:DPL,Wang:2023:REC}.
We then perform manual annotations, where two annotators (paper authors) independently label each factual claim as check-worthy or not. In cases of disagreement, the third annotator independently breaks ties to determine the final aggregated \textit{gold} labels.

\medskip \noindent \textbf{Automatic Check-Worthiness Classification.}
We denote the check-worthi\-ness classification method as $CW$,  which outputs a binary label $l \in \{0, 1\}$ for a given factual claim $a$, where $l=1$ indicates that $a$ is check-worthy and $l=0$ otherwise, i.e., $l = CW(a, h)$. 
Here, $h$ is the conversation history up to the turn where $a$ is generated, the same as in factual claim extraction in Section~\ref{sec:claim_ext:experiments}.

We use two existing check-worthiness detection approaches by Majer et al.~\cite{Majer:2024:CCD} and Hassan et al.~\cite{Hassan:2015:DCF}, referred to as $CW_{Majer}$ and $CW_{Hassan}$, respectively.
These are selected based on their simplicity, effectiveness, and representativeness from existing methods~\cite{Hassan:2015:DCF,Majer:2024:CCD,Wright:2020:CCD,Deng:2024:DCE,hassan2017toward}.
The former, $CW_{Majer}$, uses optimized prompts to identify factual and check-worthy claims, while the latter, $CW_{Hassan}$, employs simple prompting based on early check-worthiness crowdsourcing studies~\cite{Hassan:2015:DCF}; both methods with detailed documentation are available in our resource.
We also add the intersection and union of the two methods, denoted as $CW_{Intersection}(a,h) = CW_{Majer}(a,h) \land CW_{Hassan}(a,h)$ and $CW_{Union}(a,h) = CW_{Majer}(a,h) \lor CW_{Hassan}(a,h)$, respectively.
A check-worthy utterance is defined as utterance $u_{cw}$ such that $\exists a \in F(u_{cw}, h)$ where $CW(a, h) = 1$. Similarly, a check-worthy conversation is a conversation $c = \{u_{s1}, u_{s2}, \ldots, u_{sm}\}$ such that $\exists u_{cw} \in c$ where $u_{cw}$ is a check-worthy utterance.


\medskip \noindent \textbf{Automatic Check-Worthiness Setup.}
Two check-worthiness classification methods are applied to all factual claims extracted in Section~\ref{sec:claim_ext:experiments}.
In the same way as in factual claim extraction, we use GPT-4.1 for both methods' backbone model and perform minor adjustments to the prompts to accommodate conversation history, thereby allowing for accurate classification of check-worthiness of factual claims considering the whole conversation context.
All prompts used are available in our repository.

\subsection{Dataset Statistics}
The dataset statistics are shown in the left half of Table~\ref{tab:wildclaims_dataset_statistics}.
Our dataset contains 121,905 factual claims from 7,587 system utterances.
The creation of this dataset required processing a total of 988 million tokens.
This large-scale annotation enables analysis of users' access to factual information in real conversations.

The task classification statistics is shown in the right half of Table~\ref{tab:wildclaims_dataset_statistics}.
While explicit information seeking utterances are still relatively common (33.5\%), we observe that a significant proportion of utterances (66.5\%) are not explicitly information seeking, such as creative writing and editing tasks.
\section{Results \& Analysis}
\label{sec:results}

Enabled by the \dataset dataset, we are now able to analyze how users' access to factual information happens in real user-system conversations.
In this section, we present the results and analysis from our dataset, addressing the \textit{RQ2: How prevalent is conversational information access in real user-system interactions?}


\begin{table}[t]
\caption{Statistics of factual claim extraction methods applied to 3,000 conversations in \dataset.
    Utt./conv. with facts (\%) represents the percentage of utterances or conversations that contain at least one extracted factual claim.}
\label{tab:fact_extraction_stats}
\centering
\small
\setlength{\tabcolsep}{3pt}
\begin{tabular}{l|c|cc|cc}
\hline
\textbf{\makecell[l]{Claim ext.\\method}} & \textbf{\makecell[c]{\# Total\\facts}} & \textbf{\makecell[c]{Avg. facts\\per utt.}} & \textbf{\makecell[c]{Avg. facts\\per conv.}} & \textbf{\makecell[c]{Utt. with\\facts (\%)}} & \textbf{\makecell[c]{Conv. with\\facts (\%)}} \\
\Xhline{2pt}
$F_{Huo}$ & 31{,}108 & 4.1 & 10.4 & 45.6\% & 45.1\% \\
$F_{Song}$ & 90{,}797 & 12.0 & 30.3 & 72.1\% & 79.0\% \\
\hline
\end{tabular}
\end{table}

\subsection{Factual Claim Extraction}

We first present the results of the extracted factual claims; see Table~\ref{tab:fact_extraction_stats}.
We see that $F_{Song}$ extracts substantially more factual claims than $F_{Huo}$, nearly three times as many in total (90,797 vs. 31,108).
On average, $F_{Song}$ identifies 30.3 facts per conversation, compared to 10.4 for $F_{Huo}$.
Consequently, a much higher percentage of conversations contain at least one factual claim when processed by $F_{Song}$ (79.0\%) compared to $F_{Huo}$ (45.1\%).
This indicates that $F_{Song}$ provides a more comprehensive coverage of factual claims within conversations.

\subsection{Manual Check-Worthiness}
\label{sec:claim_ext}

This section presents manual check-worthiness verification results and a conservative, lower-bound estimate of check-worthy claim prevalence in conversations.

\begin{table}[t]
\caption{Statistics of human annotations for validating check-worthiness.
    Hum. 1 and Hum. 2 are two independent human annotators.
    Aggr. is the aggregated (gold) label from both annotators; ties are broken by a third annotator.
    For each claim extraction method, 100 factual claims are annotated by humans; see Section~\ref{sec:claim_ext:experiments} for details.}
\label{tab:human_checkworthiness_stats}
\centering
\setlength{\tabcolsep}{9.1pt}
\begin{tabular}{l|ccc|c}
\hline
\multirow{2}{*}{\textbf{\makecell[l]{Claim ext.\\method}}} &
\multicolumn{3}{c|}{\textbf{\makecell[l]{\% of check-worthy facts}}} &
\multirow{2}{*}{\textbf{\makecell[l]{Kappa between\\Hum. 1 and 2}}} \\
\cline{2-4}
 & \textbf{Hum. 1} & \textbf{Hum. 2} & \textbf{Aggr.} & \\
\Xhline{2pt}
$F_{Huo}$     & 43\% & 41\% & 40\% & 0.672 \\
$F_{Song}$    & 59\% & 63\%  & 64\% & 0.580 \\
\hline
\end{tabular}
\end{table} 

\medskip \noindent \textbf{Human Check-Worthiness Verification Results.}
Table~\ref{tab:human_checkworthiness_stats} shows the statistics of manual annotations.
In the table, we see inter-annotator agreement with a Cohen's Kappa score of 0.672 and 0.580 for $F_{Huo}$ and $F_{Song}$, respectively, indicating a moderate to substantial agreement for human annotations.
We also see that automatically extracted factual claims contain a significant portion of check-worthy factual claims, with 40\% and 64\% for aggregated gold labels for $F_{Huo}$ and $F_{Song}$, respectively.
This demonstrates that, while not perfect, automatic claim extraction methods do identify factual claims that are check-worthy, supporting their viability as reasonable tools for our further experiments to estimate the prevalence of factual assertions.

\medskip \noindent \textbf{Lower-Bound Prevalence Estimate.}
We now estimate the prevalence of conversations containing at least one check-worthy factual claim.
To obtain a conservative, lower-bound estimate, we make the pessimistic assumption that each conversation contains no more than one factual claim.
Let $x$ be the proportion of human-annotated check-worthy factual claims over all extracted claims (``\% of check-worthy facts -- Aggr.'' in Table~\ref{tab:human_checkworthiness_stats}) and $y$ be the proportion of conversations that contain at least one extracted factual claim (``Conv. with facts (\%)'' in Table~\ref{tab:fact_extraction_stats}), then we can conservatively estimate the proportion of conversations that contain at least one check-worthy factual claim as $x \times y$.
The necessity of multiplying $y$ arises because not all utterances contain extracted claims, as shown in Table~\ref{tab:fact_extraction_stats}.
Note that this estimate is quite conservative because, in reality, conversations often include several factual claims (see Table~\ref{tab:fact_extraction_stats}), and identifying even a single check-worthy statement requires retrieving and verifying information from external sources. 

Our conservative estimates indicate that for $F_{Huo}$, at least $18\%$ ($=45.1\% \times 40.0\%$) of conversations contain at least one check-worthy factual claim.
For $F_{Song}$, the estimate is $51\%$ ($=79.0\% \times 64.0\%$).
These results suggest that, even with a lower-bound conservative estimation method, a substantial portion of conversations contains implicit information access that occurs in the form of check-worthy factual claims.


In the next section, using automatic check-worthiness classification methods, we provide a less conservative estimate of the prevalence of check-worthy factual claims in conversations.

\subsection{Automatic Check-Worthiness}
\label{sec:cw}

\begin{table}[t]
\caption{Effectiveness of automatic check-worthiness (CW) classification evaluated using human annotations (100 annotations for each claim extraction method; see Section~\ref{sec:claim_ext:experiments}).}
\label{tab:cw-effectiveness}
\centering
\setlength{\tabcolsep}{8.1pt}
\begin{tabular}{l|l|cccc}
\hline
\multirow{2}{*}{\textbf{\makecell[l]{Claim ext.\\method}}} &
\multirow{2}{*}{\textbf{\makecell[l]{CW method}}} &
\multicolumn{4}{c}{\textbf{\makecell[l]{CW classification results}}} \\
\cline{3-6}
 & & \textbf{P} & \textbf{R} & \textbf{F} & \textbf{$\kappa$} \\
\Xhline{2pt}
\multirow{4}{*}{$F_{Huo}$}
    & $CW_{Hassan}$      & 0.610 & 0.900 & 0.727 & 0.479 \\
    & $CW_{Majer}$        & 0.697 & 0.575 & 0.630 & 0.421 \\
    & $CW_{Intersection}$ & \textbf{0.700} & 0.525 & 0.600 & 0.391 \\
    & $CW_{Union}$ & 0.613 & \textbf{0.950} & \textbf{0.745} & \textbf{0.504} \\
\hline
\multirow{4}{*}{$F_{Song}$}
    & $CW_{Hassan}$       & 0.763 & 0.906 & 0.829 & 0.438 \\
    & $CW_{Majer}$        & \textbf{0.780} & 0.500 & 0.610 & 0.219 \\
    & $CW_{Intersection}$ & 0.769 & 0.469 & 0.583 & 0.190 \\
    & $CW_{Union}$ & 0.769 & \textbf{0.938} & \textbf{0.845} & \textbf{0.478} \\
\hline
\end{tabular}
\end{table}

We showed the lower-bound estimate of the prevalence of check-worthy claims in real-world conversations (cf. Section~\ref{sec:claim_ext}). This section presents a less conservative estimate of the prevalence of check-worthy factual claims in conversations using existing automatic check-worthiness classification methods.


\begin{table}[t]
\caption{Prevalence of check-worthy (CW) claims in 3,000 conversations in the \dataset dataset, estimated with CW classifiers. The \% CW facts/utterances/conversations column reports the share of claims tagged CW and the share of utterances or conversations containing at least one CW claim; see Section~\ref{sec:cw:experiments}.}
\label{tab:cw-prevalence}
\centering
\small
\setlength{\tabcolsep}{4pt}
\begin{tabular}{l|l|ccc|cc}
\hline
\multirow{2}{*}{\textbf{\makecell[l]{Claim ext.\\method}}} &
\multirow{2}{*}{\textbf{\makecell[l]{CW method}}} &
\multicolumn{3}{c|}{\textbf{\makecell[l]{\% CW facts/utt/conv}}} &
\multicolumn{2}{c}{\textbf{\makecell[l]{\# CW facts\\per utt/conv}}} \\
\cline{3-7}
 & & \makecell[c]{\textbf{facts}} &
     \makecell[c]{\textbf{utt.}} &
     \makecell[c]{\textbf{conv.}} &
     \makecell[c]{\textbf{utt.}} &
     \makecell[c]{\textbf{conv.}} \\
\Xhline{2pt}
\multirow{4}{*}{$F_{Huo}$}
    & $CW_{Hassan}$       & 68.9\% & 41.0\% & 41.0\% & 2.8 & 7.1 \\
    & $CW_{Majer}$        & 33.7\% & 28.7\% & 32.4\% & 1.4 & 3.5 \\
    & $CW_{Intersection}$ & 32.9\% & 28.3\% & 31.7\% & 1.4 & 3.4 \\
    & $CW_{Union}$        & \textbf{69.6\%} & \textbf{41.1\%} & \textbf{41.3\%} & \textbf{2.9} & \textbf{7.2} \\
\hline
\multirow{4}{*}{$F_{Song}$}
    & $CW_{Hassan}$       & 91.3\% & 70.1\% & 76.2\% & 10.9 & 27.6 \\
    & $CW_{Majer}$        & 51.5\% & 58.3\% & 64.3\% & 6.2 & 15.6 \\
    & $CW_{Intersection}$ & 51.0\% & 58.0\% & 64.0\% & 6.1 & 15.4 \\
    & $CW_{Union}$        & \textbf{91.8\%} & \textbf{70.2\%} & \textbf{76.4\%} & \textbf{11.0} & \textbf{27.8} \\
\hline
\end{tabular}
\end{table}

\medskip \noindent \textbf{Automatic Check-worthiness Effectiveness.}
Table~\ref{tab:cw-effectiveness} shows the effectiveness of automatic check-worthiness classification methods evaluated using human gold annotations collected in Section~\ref{sec:claim_ext:experiments}.
Comparing $CW_{Hassan}$ and $CW_{Majer}$, we observe that while $CW_{Majer}$ has higher precision, $CW_{Hassan}$ outperforms in all other metrics, recall, F1-score, and $\kappa$.
Of the four check-worthiness classification methods, $CW_{Union}$ demonstrates the highest performance, achieving a kappa of 0.504 for the $F_{Huo}$ method and 0.478 for the $F_{Song}$ method.
This indicates moderate agreement with human gold annotations, suggesting that $CW_{Union}$ is a more reliable choice for prevalence estimation.

\medskip \noindent \textbf{Prevalence Estimation.}
Table~\ref{tab:cw-prevalence} shows the prevalence of check-worthy factual claims in the \dataset dataset using automatic check-worthiness classification methods.
At a high level, we observe that the prevalence of check-worthiness varies significantly between the two claim extraction methods. This is because, as shown in Table~\ref{tab:fact_extraction_stats}, $F_{Huo}$ extracts significantly fewer factual claims than $F_{Song}$, which results in a lower prevalence of check-worthy factual claims, suggesting many missing factual claims from $F_{Huo}$.
Focusing on the $F_{Song}$, which extracts higher percentage of check-worthy factual claims according to human annotations (see Table~\ref{tab:human_checkworthiness_stats}), and using $CW_{Union}$ as it achieves the highest correlation with human annotations (see Table~\ref{tab:cw-effectiveness}), the prevalence of check-worthy conversations, which is the percentage of conversations that contain at least one check-worthy factual claim, is 76.4\% (2,291 out of 3,000 conversations), which is strikingly high given that only 33.5\% of conversations are explicitly information-seeking (see Table~\ref{tab:wildclaims_dataset_statistics}).
This finding reinforces the notion that check-worthy factual claims are prevalent in real-world conversations, even when the conversations are not explicitly information-seeking.

\begin{figure}[t]
    \centering
    \includegraphics[width=\linewidth]{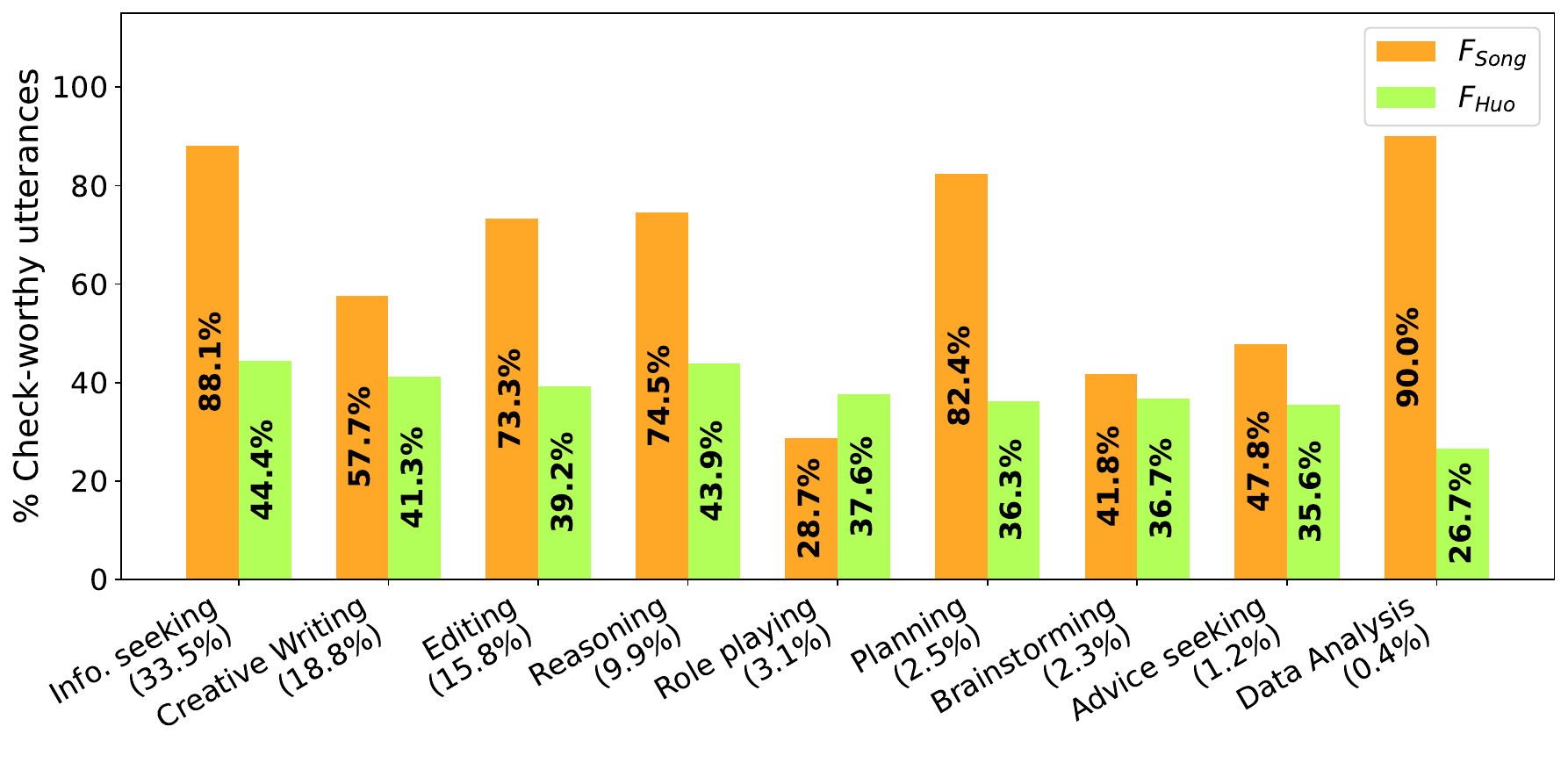}
    \caption{Prevalence of check-worthy utterances in the \dataset dataset by task category, using $CW_{Union}$ as the classification method due to its highest correlation with human annotations (see Table~\ref{tab:cw-effectiveness}).}
    \label{fig:prevalence}
\end{figure}

\medskip \noindent \textbf{Prevalence of Check-Worthy Utterances by Task Category.}
Figure~\ref{fig:prevalence} shows the prevalence of check-worthy utterances in 3,000 conversations from \dataset, broken down by task categories.
We see check-worthy utterances are prevalent across all task categories, e.g., in creative writing, 57.7\% and 41.3\% of utterances are check-worthy according to $F_{Song}$ and $F_{Huo}$, respectively.

This further supports our observation discussed in Section~\ref{sec:wild:dataset} that information access in real-world conversations happens across a variety of contexts, not restricted to explicit information-seeking intent, and even when the user intent is seemingly unrelated to information seeking or recommendation.

\section{Conclusion}

This paper examined information access in real-world user-system conversations.
Our observational analysis on user-ChatGPT conversations revealed that users' information access to factual information occurs implicitly through factual claims made by the system, even during seemingly non-information-access tasks like creative writing or editing.
To enable the study of this phenomenon, we developed the \textbf{\dataset} dataset, a novel resource of 121,905 factual claims from 7,587 system utterances across 3,000 conversations, each annotated for check-worthiness.
With this dataset, we estimate conservatively that 18\%--51\% of conversations contain factual assertions requiring verification.
This underscores the need to move beyond explicit information access to the implicit knowledge transfer prevalent in real-world conversations.
We emphasize that our work is a preliminary attempt to understand the scope of the problem, rather than to propose a final solution.
Our annotated dataset (\dataset: \href{https://github.com/shakibaam/wildclaims}{https://github.com/shakibaam/wildclaims}) opens up new opportunities for future research, including developing more precise tools to identify which factual claims require verification.

\bibliographystyle{splncs04}
\bibliography{references-base}

\end{document}